# An Artificial Intelligence Model for Early-stage Breast cancer detection from Biopsy images

N. Chaudhary[a] & A.Z. Dhunny[b]

[a] *Pangea Society, India*
[b] *Artificial Intelligence Engineer, Cyber Analytics, Mauritius*

**Abstract**

Accurate identification of breast cancer types plays a critical role in guiding treatment decisions and improving patient outcomes. This paper presents an artificial intelligence enabled tool designed to aid in the identification of breast cancer types using histopathological biopsy images. Traditionally additional tests have to be done on women who are detected with breast cancer to find out the types of cancer it is to give the necessary cure. Those tests are not only invasive but also delay the initiation of treatment and increase patient burden. The proposed model utilizes a convolutional neural network (CNN) architecture to distinguish between benign and malignant tissues as well as accurate subclassification of breast cancer types. By preprocessing the images to reduce noise and enhance features, the model achieves reliable levels of classification performance. Experimental results on such datasets demonstrate the model's effectiveness, outperforming several existing solutions in terms of accuracy, precision, recall, and F1-score. The study emphasizes the potential of deep learning techniques in clinical diagnostics and offers a promising tool to assist pathologists in breast cancer classification.

**Keywords:** Breast Cancer Detection; Artificial Intelligence; Convolutional Neural Network; Sub-Classes of Breast Cancer.

# 1.Introduction

Breast cancer remains one of the most prevalent cancers globally, representing a significant health challenge for women across all age groups. According to the World Health Organization (WHO), over 2.3 million women were diagnosed with breast cancer in 2020, making it the most diagnosed cancer worldwide and the leading cause of cancer-related deaths among women (WHO, 2021). The incidence of breast cancer is rising by around 3% per year, with higher mortality rates observed in lower-income countries due to limited access to early screening and treatment. In wealthier nations, 1 in 12 women are diagnosed with breast cancer, whereas in lower-income countries, the rate is 1 in 27. More concerning is the disparity in mortality—1 in 48 women die from breast cancer in low-income countries compared to 1 in 71 in high-income countries (WHO, 2022). In sub-Saharan Africa, breast cancer now has the highest mortality rate among all cancers affecting women, surpassing cervical cancer. It accounts for 20% of cancer-related deaths in women, with incidence rates varying by region: 30.4 per 100,000 women in Eastern Africa, 26.8 in Central Africa, 38.6 in Western Africa, and 38.9 in Southern Africa. Despite lower incidence rates than in developed countries, the mortality-to-incidence ratio remains alarmingly high at 0.55 in Central Africa, compared to just 0.16 in the United States (GLOBOCAN, 2020).

Early detection has been identified as a critical factor in improving survival rates, with studies showing that early-stage breast cancer has a 90% five-year survival rate compared to late-stage diagnoses, which can drop to 27% (Siegel et al., 2022) . This significant effort to enhance diagnostic techniques to detect breast cancer earlier and more accurately. Over the past decade, there have been remarkable advances in breast cancer detection, particularly with the integration of advanced imaging techniques and machine learning. Traditional imaging modalities, such as mammography and ultrasound, have been instrumental in identifying breast lesions; however, their sensitivity can decrease in women with dense breast tissue (Kolb et al., 2002). Magnetic resonance (MRI) has emerged as a superior modality for detecting early-stage breast cancer due to its high spatial resolution and ability to differentiate between benign and malignant lesions (Kuhl et al., 2005). Despite its advantages, retention remains complex and prone to variability among radiologists, emphasizing the need for more standardized and accurate approaches. Artificial Intelligence models present a transformative solution to these challenges, especially in resource-limited settings. By leveraging AI to analyze biopsy images, diagnostic accuracy can be improved while reducing the workload of overburdened pathologists. AI-powered systems have demonstrated sensitivities of up to 94.5% in detecting malignant tumors in histopathological images, reducing false-negative rates significantly (Esteva et al., 2021). Additionally, AI-assisted diagnostics have the potential to cut down diagnosis time from weeks to just hours, allowing for earlier treatment initiation and improved patient outcomes. This is particularly beneficial in low-resource settings where a shortage of trained pathologists delays cancer detection (McKinney et al., 2020). By integrating AI into breast cancer diagnostics, healthcare systems in developing countries can bridge the gap in early detection, reduce misdiagnosis rates, and ultimately lower breast cancer mortality. Scalable AI solutions, combined with improved screening programs and public awareness efforts, have the potential to significantly enhance cancer care worldwide.

## 1.1 Brief review of Artificial intelligence in Breast cancer

This section provides a brief literature review of what technology has been used for the detection of breast cancer over the past few years for breast cancer detection. And this is going to justify our innovative model and method in further sections.

AI has revolutionized the healthcare landscape, offering transformative capabilities in automating and enhancing diagnostic processes. In the context of breast cancer detection, AI models have demonstrated superior performance in analyzing complex MRI datasets, identifying patterns that may elude human experts (Lehman et al., 2019). For instance, deep learning convolutional neural networks (CNNs) and attention mechanisms have been applied to segment breast tissues, classify lesions, and predict malignancy with high accuracy (Litjens et al., 2017) . These advancements not only reduce errors but also minimize the workload of radiologists, making AI a valuable tool in clinical settings. The integration of AI into breast cancer detection workflows has also been shown to address the challenges of class imbalance and variability in imaging quality. Many studies have employed transfer learning and ensemble techniques to overcome these challenges, achieving higher sensitivity and specificity compared to traditional methods (McKinney et al., 2020) . Moreover, AI-driven algorithms can process datasets quickly and consistently, offering significant advantages in population-based breast cancer screening programs (Rodrigues et al., 2021) .

Wang et al., (2020) leveraged the Northwestern Medicine Enterprise Warehouse dataset to predict breast cancer distant recurrence by integrating structured clinical data with unstructured text processed into word embeddings and UMLS Concept Unique Identifiers (CUIs). Using models like Random Forests, SVMs, and a knowledge-guided CNN, the study demonstrated the effectiveness of combining NLP and machine learning techniques to enhance predictive accuracy and support personalized healthcare interventions. Souza et al. (2023) utilized ATR-FTIR spectroscopy on blood plasma samples to discriminate molecular subtypes of breast cancer (Luminal A, Luminal B, HER2, and Triple-negative). The study combined partial least square-artificial neural network discriminant analysis (PLS-ANNDA) for accurate subtype classification, leveraging spectral patterns to identify biochemical differences. This non-invasive approach highlights the potential of spectroscopy and advanced chemometric techniques for early detection and personalized cancer management. Sertan Kaymak et al. (2017) explored breast cancer classification using artificial neural networks trained on 176 histopathology images (112 cancerous, 64 non-cancerous) from Near East University Hospital. Images were processed with discrete Haar wavelets to enhance edge clarity before being classified using Back Propagation Neural Networks (BPNN) and Radial Basis Function Networks (RBFN). The RBFN outperformed BPNN with a 70.49% accuracy. This study highlights the role of advanced image processing and neural network models in enhancing breast cancer detection in histopathology images.  Fu et al. (2022) developed a predictive model for PICC-related thrombosis in breast cancer patients undergoing chemotherapy, using a prospective cohort of 1,844 patients. Variables such as age, comorbidities, and catheter details were selected via regression analysis, and SMOTE addressed class imbalance. An artificial neural network (ANN) achieved superior performance (AUC 0.742) compared to logistic regression (AUC 0.675). This study demonstrates the potential of machine learning in enhancing risk prediction and improving clinical outcomes for breast cancer patients with PICCs. Punitha S. et al. (2021) leveraged the Wisconsin Breast Cancer Dataset (WBCD) to develop an automated diagnosis system using Artificial Neural Networks (ANN). The study introduced a novel approach combining Artificial Immune Systems and Artificial Bee Colony algorithms for feature selection, alongside simulated annealing for parameter optimization. Utilizing a Multilayer Perceptron (MLP) model, this method improved diagnostic accuracy by optimizing network architecture and learning parameters, offering a robust and efficient tool for automated breast cancer diagnosis. Sharma et al. (2024) utilized the Wisconsin Breast Cancer Dataset (WBCD) to develop a stacked ensemble framework for breast cancer prediction. By integrating models like Decision Tree, AdaBoost, Gaussian Naive Bayes, and Multilayer Perceptron (MLP), the framework achieved a high accuracy of 97.66%. Advanced ensemble techniques and extensive feature engineering ensured robust performance, demonstrating the potential of such approaches to enhance diagnostic precision and support clinical decision-making.  Liu et al. (2024a) developed an improved VGG16-based method for classifying breast cancer from a dataset of 7,841 mammography images from Ganzhou People's Hospital and The Sixth Affiliated Hospital of Jinan University. Enhanced with transfer learning and a focal loss function to address class imbalance, the model achieved a classification accuracy of 96.945%. This study underscores the efficacy of tailored deep learning techniques in medical imaging and their potential for robust breast cancer diagnostics.  Li et al. (2024b) developed a multimodal fusion model integrating MRI and RNA-seq data to predict pathological responses to neoadjuvant chemotherapy in 165 breast cancer patients. MRI features were extracted using ResNet34, and RNA-seq data focused on differentially expressed genes, with an attention mechanism and transformer techniques enhancing fusion and interpretation. This approach outperformed single-modality models, demonstrating the potential of deep learning in integrating diverse medical data for personalized treatment strategies.

Miguel et al. (2023) explored the use of evolutionary algorithms for training neural networks to classify histological breast cancer images from Massachusetts General Hospital and Beth Israel Deaconess Medical Center. The dataset consisted of 343 stained images, categorized into ductal carcinoma in situ (DCIS) and usual ductal hyperplasia (UDH). Feature extraction included color, texture (using Local Binary Patterns), and CNN-based features from ResNet50. Evolutionary algorithms like Genetic Algorithm, Differential Evolution, and Particle Swarm Optimization were used to optimize the neural network, avoiding local minima. The study demonstrated the effectiveness of evolutionary methods, with Differential Evolution achieving an AUC of 0.70 using color features and PSO scoring 0.71 with texture features, showing promise for improving histological image classification in breast cancer diagnosis. Guo et al. (2024) proposed a multimodal breast cancer diagnosis method using a proprietary video dataset with 332 cases of Contrast-Enhanced Ultrasound (CEUS) and B-mode Ultrasound (B-US). CEUS provided dynamic vascular flow, while B-US offered structural data. The KAMnet (Knowledge-Augmented Network) framework incorporated medical knowledge to enhance model interpretability and performance. Temporal sampling focused on keyframes during contrast agent wash-in and wash-out phases, and a feature fusion network combined spatial (B-US) and temporal (CEUS) information. The model achieved a sensitivity of 90.91% and accuracy of 88.24%, with ablation studies highlighting the effectiveness of knowledge-augmented modules in improving prediction accuracy and handling limited data. Munshi et al. (2024) introduced a novel breast cancer detection method by combining an optimized ensemble learning framework with Explainable AI (XAI). Using the Wisconsin Breast Cancer Dataset, which contains 32 numerical features, they employed an ensemble model integrating Convolutional Neural Networks (CNN) with traditional machine learning algorithms like Random Forest (RF) and Support Vector Machine (SVM). U-NET was used for image-based tasks, and a voting mechanism combined RF and SVM predictions. The model achieved an impressive accuracy of 99.99%, surpassing state-of-the-art methods in precision, recall, and F1 score. The integration of XAI enhanced the interpretability and transparency of model decisions, further improving breast cancer diagnostics.

Atrey et al. (2024) developed a multimodal classification system for breast cancer by integrating mammogram (MG) and ultrasound (US) images using deep learning and traditional machine learning techniques. The dataset, sourced from AIIMS, Raipur, India, consisted of 31 patients and 43 MG and 43 US images, which were augmented to 1,032 images. Data preprocessing involved manual region-of-interest (ROI) annotation by radiologists and noise filtering techniques for both image types. The deep learning model, ResNet-18, was used for feature extraction, while Support Vector Machine (SVM) classified the fused features. The hybrid ResNet-SVM approach achieved a classification accuracy of 99.22%, significantly outperforming unimodal methods. This study demonstrates the effectiveness of combining deep learning with traditional machine learning for improved breast cancer diagnosis. G. Robinson Paul and J. Preethi (2024) developed a breast cancer detection system using advanced segmentation and recurrent neural networks (RNN). The study used 9,109 histopathological images from the BreaKHis and BH datasets (2,480 benign, 5,429 malignant) at various magnifications. Preprocessing included resizing to 512x512 pixels, noise reduction, and stain normalization. The SDM-WHO-RNN model combined a fully convolutional network with RNN for classification, while Linear Scaling-Canny Edge Detection (LS-CED) was used for precise tumor segmentation. The system achieved a classification accuracy of 97.9%, improving detection of complex tumor cells. Zhang et al. (2021) developed a novel Voting Convergent Difference Neural Network (V-CDNN) to enhance breast cancer diagnosis. The approach utilized publicly available breast cancer datasets, applying feature selection and normalization for data preprocessing. The V-CDNN combined multiple Convergent Difference Neural Networks (CDNNs) using ensemble learning with a plurality voting strategy for improved accuracy. The model employed a Neural Dynamic Algorithm (NDA) for efficient error convergence and used softsign activation functions to accelerate training. The system achieved a perfect classification accuracy of 100%, offering superior efficiency and rapid training compared to traditional methods.

In their 2021 study, Meha Desai and Manan Shah compared the effectiveness of Multi-Layer Perceptron (MLP) and Convolutional Neural Network (CNN) for breast cancer detection. The study used datasets such as BreakHis, WBCD, and WDBC, with images and biomarkers preprocessed through normalization and feature extraction (e.g.,

Discrete Cosine Transform). While both MLP and CNN were trained to classify breast abnormalities, CNN consistently outperformed MLP in terms of accuracy. CNN achieved ~99.86% accuracy on the BreakHis dataset, while MLP's performance was generally lower. CNN was found to be more reliable for image-based diagnosis, whereas MLP showed limitations for larger datasets. In the 2024 study by Akshata K. Naik and Venkatanareshbabu Kuppili, a Weighted Generalized Classifier Neural Network (WGCNN) was proposed for feature selection in microarray gene expression datasets. These datasets are known for their high dimensionality and noise. Preprocessing involved statistical-guided dropout to avoid overfitting. The WGCNN incorporated embedded feature weighting, allowing it to capture non-linear feature interactions, and its five-layer architecture ensured model explainability. Compared to sparse group Lasso and other regularization techniques, WGCNN demonstrated improved F1 scores and reduced feature dimensions without sacrificing classification accuracy (Desai & Shah, 2021). It's an explainable nature which provided an advantage over deep learning's often opaque models. Further to this, Ezzat et al. (2023) proposed an optimized Bayesian convolutional neural network (OBCNN) for detecting invasive ductal carcinoma (IDC) from histopathology images. This approach integrated ResNet101V2 with Monte Carlo dropout (MC-dropout) for uncertainty estimation. Using the slime mould algorithm (SMA) to optimize dropout rates, the study fine-tuned the architecture, outperforming other pre-trained networks such as VGG16 and DenseNet121. The model demonstrated significant robustness in diagnosis and generalization. Farajzadeh et al. (2020) introduced a residual fully convolutional encoder-decoder network for localizing cancer nuclei in histopathology images. By masking cancer nuclei and enabling segmentation at the pixel level, their approach facilitated accurate detection of cancerous regions. The model employed Dice Loss as an evaluation metric to ensure reliable results, overcoming the limitations of less accurate methods. Joseph et al. (2022) proposed a multi-classification approach for breast cancer using handcrafted features, including Hu moments, Haralick textures, and color histograms. These features were combined with deep neural networks trained on the BreakHis dataset. Data augmentation was applied to address overfitting, and a four-layer dense network with Softmax activation was employed to enhance classification accuracy. Zhang et al. (2020) developed a hybrid model called BDR-CNN-GCN, which combined graph convolutional networks (GCN) and convolutional neural networks (CNN). This method utilized batch normalization and dropout to improve robustness while employing rank-based stochastic pooling to replace traditional max pooling. The model achieved enhanced performance in classifying breast mammograms. Paul et al. (2023) proposed a novel breast cancer detection system using an SDM-WHO-RNN classifier combined with LS-CED segmentation. Their approach incorporated preprocessing steps such as noise elimination and image normalization. The LS-CED segmentation localized nuclei, which were then classified based on size and shape. The system demonstrated high accuracy in detecting cancerous cells. M. Taheri and H. Omranpour (2023) introduced an ensemble meta-feature generator (EMFSG-Net) for classifying ultrasound images. Leveraging transfer learning with the VGG-16 architecture, the model employed Support Vector Regression (SVR) to create efficient feature spaces. To address issues like overfitting and dead neurons, the Leaky-ReLU activation function was incorporated, leading to improved feature representation and classification performance. Muduli et al. (2021) proposed a deep convolutional neural network for automated breast cancer classification using mammograms and ultrasound images. The model, consisting of five learnable layers, employed manual cropping for feature extraction and data augmentation to improve generalization. The approach achieved superior results compared to state-of-the-art methods. Kaymak et al. (2018) utilized artificial neural networks for breast cancer image classification. By applying Gaussian filters and discrete Haar wavelets to preprocess images, their method reduced background noise and enhanced edge clarity. A feed-forward neural network with backpropagation, further refined with Radial Basis Function Networks (RBFN), significantly improved classification performance.

This research is towards an innovative approach for breast cancer detection including the multiple subclasses of breast cancer, which still confuses medical doctors and hence making patients undergoing more advanced invasive tests for more understanding of the disease. The technique is based on an Artificial intelligence model trained with a sizable batch of data. Section 2 discusses the model in details; section 3 shows the results & discussion and the conclusion is given in section 4.

## 2. Methodology

This methodology section combines an explanation on the different types of breast cancer and their subclasses along with an advanced Artificial Intelligence technique which has been developed to analyse breast cancer results of biopsy. The aim is instead of making women go again and again through trial and errors with treatments, or misdiagnose the cancer, using AI can make all the difference. As mentioned in the literature review section above, AI tech has changed the way the world views medicine now. Therefore, in this study, we have designed a powerful tool for diagnosing the multiple subclasses also. This goes towards a humanitarian cause as women are the target here.

Benign and malignant breast cancers differ in their behavior, prognosis, and treatment approach. Benign tumors, such as fibroadenomas or cysts, are non-cancerous growths that do not invade surrounding tissues or spread to other parts of the body. They tend to have well-defined borders, grow slowly, and usually pose little to no health risk. In contrast, malignant tumors are cancerous and have the potential to invade nearby tissues and metastasize to distant organs through the lymphatic system or bloodstream (Robbins et al., 2010). Malignant breast cancers, such as Invasive Ductal Carcinomas , Invasive Lobular Carcinomas etc show uncontrolled cell growth and can require serious medical intervention in the form of surgery and/or radiation therapy. Early identification of a tumour as benign or malignant is critical to deliver appropriate treatment and improving patient health (National Cancer Institute, 2021). In histopathological images, benign and malignant tissues exhibit discernibly unique optical features. Benign tumours present as well organised structures consisting of cells with uniform shapes, minimal mitotic activity and intact basement membranes. The stromal and glandular components of benign lesions are usually preserved, showing regular nuclei and minimal pleomorphism. On the other hand, malignant lesions show irregular cellular arrangements, pleomorphic nuclei and frequent mitotic figures (Robbins et al., 2010). They often display disrupted basement membranes and an increased frequency of necrotic regions further highlighting their aggressive nature. AI image analysis of biopsy images will be able to use these intrinsic differences to ensure earlier and more reliable classification into benign and malignant cases (Rakhlin et al., 2018).

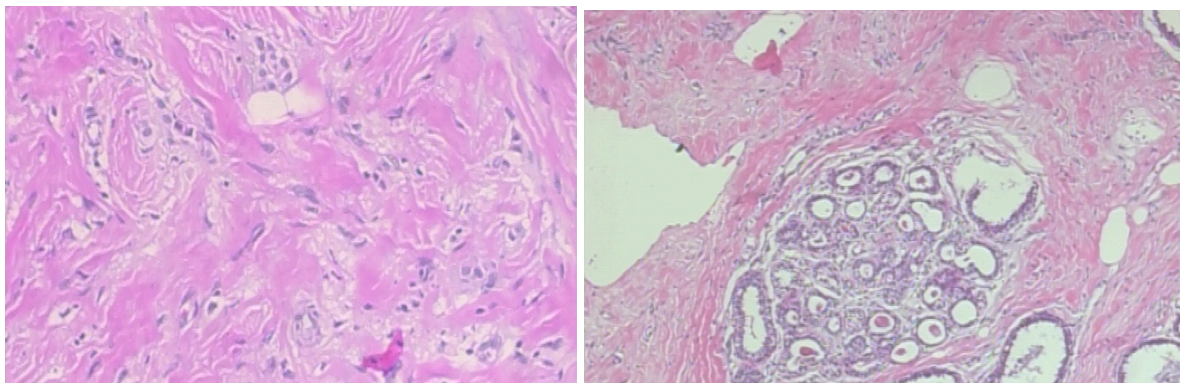

      **(a)Malignant Sample**      **(b)Benign Sample**

**Fig 1:** Biopsy Image of Malignant and Benign Sample

The visual features in the biopsy images provide key characteristics for distinguishing malignant and benign breast tissues. Figure 1 (a) (malignant sample) exhibits irregular, disorganized cell structures with pleomorphic nuclei and a dense, fibrous stroma, indicating uncontrolled cancerous growth. The nuclei appear darker and more varied in shape, with increased mitotic activity and loss of structural integrity (Robbins et al., 2010). In contrast, Figure 1 (b) (benign sample) shows well-organized, rounded glandular structures surrounded by normal fibrous tissue, with uniform cell shapes and minimal pleomorphism (National Cancer Institute, 2021).

**2.1 Dataset**

In this study, we utilize the BreaKHis (Breast Cancer Histopathological Image) dataset compiled by Spanhol et al. (2016), which contains 9,109 microscopic images of breast tissue captured at four magnification levels: 40x, 100x, 200x, and 400x. These images are categorized into two overarching classes—benign and malignant tumors. Within each class, the dataset further identifies four subtypes: for benign tumors, these include adenosis (A), fibroadenoma (F), phyllodes tumor (PT), and tubular adenoma (TA); and for malignant tumors, ductal carcinoma (DC), lobular carcinoma (LC), mucinous carcinoma (MC), and papillary carcinoma (PC). The image filenames in the BreaKHis dataset encode multiple details about each sample, including the biopsy method used, the tumor's classification (benign or malignant), its histological subtype, the patient ID, and the magnification level. For instance, the file named SOB_B_TA-14-4659-40-001.png refers to the first image of a benign tumor classified as tubular adenoma, obtained from patient 14-4659 at a magnification of 40x using the SOB (Segmental Orthogonal Biopsy) technique (Spanhol et al., 2016).

The dataset comprises samples from 82 individuals—24 with benign tumors and 58 with malignant tumors. Of the 7,909 images used in the final analysis, 5,429 are malignant and 2,480 are benign, distributed across the aforementioned magnification levels. Following preprocessing and class balancing, the revised class distribution is outlined in Table 1.

Benign tumors, by histological standards, lack features associated with malignancy such as cellular atypia, mitotic activity, invasive growth, or metastatic potential. They typically exhibit localized, slow growth and are considered non-lethal (Robbins & Cotran, 2010). In contrast, malignant tumors are characterized by their capacity to invade neighboring tissues and metastasize to distant sites, often posing life-threatening risks (Robbins & Cotran, 2010). All tissue samples in this dataset were acquired through Segmental Orthogonal Biopsy (SOB)—a surgical procedure also referred to as partial mastectomy or excisional biopsy. Unlike needle-based techniques, SOB yields larger tissue specimens and is generally performed under general anesthesia in a clinical setting (American Cancer Society, 2019). The classification of tumors into distinct subtypes is based on their cellular morphology under microscopic examination, which holds significance for both prognosis and treatment planning (National Cancer Institute, 2021).

| Cancer Type | Final Image Count | Sub Type |
|---|---|---|
| Adenosis | 444 | Benign |
| Fibroadenoma | 730 | Benign |
| Phyllodes Tumor | 453 | Benign |
| Tubular Adenoma | 569 | Benign |
| Ductal Carcinoma | 797 | Malignant |
| Lobular Carcinoma | 626 | Malignant |
| Mucinous Carcinoma | 792 | Malignant |
| Papillary Carcinoma | 560 | Malignant |

Table 1: Dataset Image Count for Each Cancer Type

## 2.2 Artificial intelligence model

A Convolutional Neural Network (CNN) is a type of deep learning model designed for structured grid-like data, such as images. CNNs extract and learn hierarchical features using convolutional layers, activation functions, pooling layers, and fully connected layers. Unlike traditional artificial neural networks (ANNs), CNNs take advantage of spatial hierarchies to capture local and global patterns, improving performance in tasks like classification, object detection, and segmentation (LeCun et al., 1998). As illustrated in Figure 2, a CNN begins with an input image that is passed through a series of convolutional layers, each applying learnable filters (kernels) to extract spatial features such as edges, textures, and shapes. Each filter slides across the input image, computing the dot product between the filter's weights and the local receptive field, producing a feature map (Krizhevsky et al., 2012). These feature maps retain spatial relationships, making CNNs highly effective at recognizing patterns regardless of their position in the image.

Non-linearity is introduced using activation functions, typically ReLU (Rectified Linear Unit), which allows the network to learn complex patterns by breaking linearity (Glorot et al., 2011). To reduce computational complexity and enhance feature invariance, pooling layers, such as max pooling, downsample feature maps by selecting the most significant values in small regions, thereby preserving essential features while reducing spatial dimensions (Scherer et al., 2010). These steps are visually represented in Figure 2, showing how the data flows from the convolutional and ReLU layers to the pooling layer. Deeper layers of the CNN capture higher-level abstractions, such as object parts and entire structures. After multiple convolutional and pooling layers, the extracted features are flattened and fed into a fully connected layer (FC layer), which performs classification based on learned representations (Simonyan & Zisserman, 2014). Finally, the output layer, often utilizing the softmax function, assigns probabilities to different classes in a classification task. CNNs have significantly outperformed traditional methods in image analysis, demonstrating state-of-the-art performance in benchmarks such as ImageNet (Russakovsky et al., 2015). Advancements such as batch normalization, residual connections (ResNets), and attention mechanisms have further improved CNN architectures for deep feature learning (He et al., 2016). These models continue to be widely used in medical imaging, autonomous driving, and other AI-driven fields.

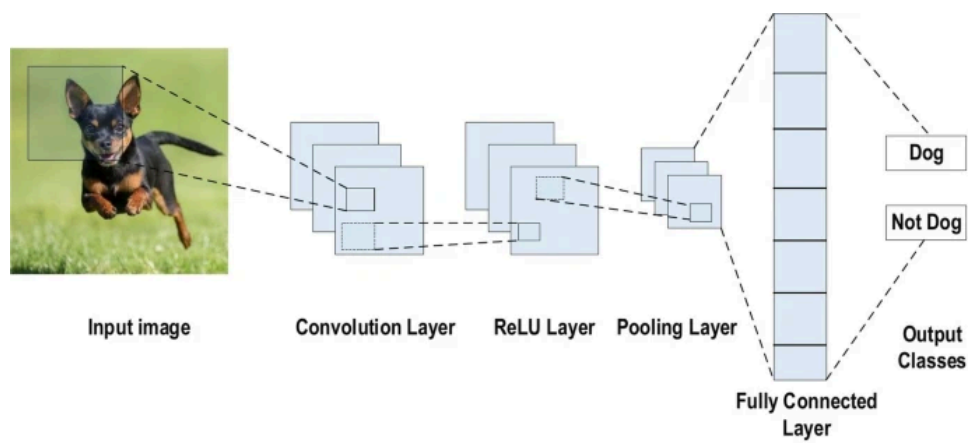

**Fig 2:** CNN architecture, Alzubaidi, L., Zhang, J., Humaidi, A.J. *et al.* (2021)

### 2.2.1 Detecting the two Categories of cancer-Malignant & Bening

In the first approach, a standard CNN was implemented to classify histopathological breast cancer images into benign and malignant categories. The parameters used in the initial approach were not arbitrarily selected but were determined after extensive testing of different configurations, as outlined in the following sections, which presented the corresponding outcomes for each variation. Various filter sizes and mesh layer configuration were experimented with to find an optimal balance for highest accuracy. Along with this experimentation, parameters were also chosen with reference to usage in previous computational works in medicinal image classification.

The architecture consisted of three convolutional layers with 16, 32, and 16 filters, allowing the network to learn hierarchical spatial features from the dataset. Research suggests that multiple convolutional layers with varying filter sizes improve feature extraction for histopathological images (Spanhol et al., 2016). A 4×4 kernel size was chosen to capture broader patterns in the images while maintaining computational efficiency, as kernel sizes between 3×3 and 5×5 are commonly used for medical image classification (Bayramoglu et al., 2017). ReLU activation functions were used in each convolutional layer to introduce non-linearity, preventing the vanishing gradient problem, a widely adopted technique in CNN-based biomedical imaging models (Litjens et al., 2017). To reduce spatial dimensions while preserving important features, MaxPooling layers were added after each convolutional block, a practice shown to enhance classification accuracy in breast cancer histopathology (Rakhlin et al., 2018). The fully connected layer contained 256 neurons, ensuring a dense representation of learned features before classification, as previous studies have found that fully connected layers with 128–512 neurons optimize model complexity and performance in similar tasks (Arevalo et al., 2016). The output layer used a sigmoid activation function, converting predictions into probability values for binary classification, a standard approach in medical image classification (Gandomkar et al., 2018). Binary crossentropy was selected as the loss function due to its effectiveness in handling binary classification tasks, ensuring that predicted probabilities aligned well with actual labels (Liu et al., 2019). The Adam optimizer, with an initial learning rate of 0.0001, was used for training, as it adaptively adjusts learning rates for stable convergence and has been shown to perform well in histopathological image classification (Bandi et al., 2018). The model was trained for 20 epochs, a typical range for CNN-based medical image classification tasks where limited data is available (Rakhlin et al., 2018). However, this approach suffered from class imbalance, as there were significantly more malignant samples than benign ones, causing the model to bias towards malignancy. Furthermore, all images were treated as part of a single dataset, without considering variations in magnification levels (40X, 100X, 200X, and 400X), leading to inconsistent feature representations and a poor accuracy, f1 score, recall, and weak validation performance. Figure 3 below shows a diagrammatic representation of the CNN architecture, with the red blocks representing the convolutional layers and the blue blocks representing the flattening layers.

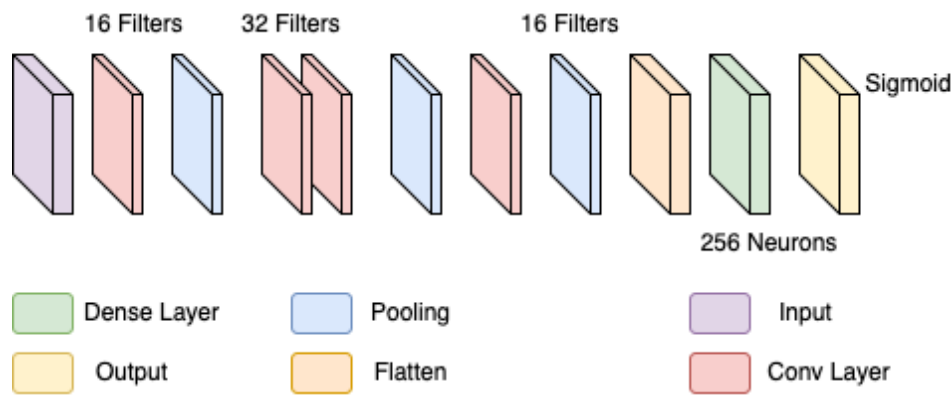

**Fig 3:** Initial CNN

| Hyperparameter | Value | Explanation |
| --- | --- | --- |
| Convolutional Filters | 16, 32, 16 | The number of filters in each convolutional layer. More filters allow for learning more complex features but increase computational cost. |
| Kernel Size | 4x4 | Defines the size of the filter that scans the image. A smaller kernel captures fine details, while a larger one captures broader patterns. |
| Activation Function | ReLU | Introduces non-linearity to the network, allowing it to learn complex patterns. ReLU prevents the vanishing gradient problem. |
| Pooling Type | MaxPooling | Reduces spatial dimensions, retaining the most important features while lowering computational cost and reducing overfitting. |
| Fully Connected Layer | 256 neurons | Processes extracted features and maps them to the final classification decision. A higher number of neurons allows for better learning capacity. |
| Output Activation | Sigmoid | Outputs a probability score between 0 and 1, making it suitable for binary classification tasks. |
| Loss Function | Binary Crossentropy | Measures how well the predicted probabilities match the true labels, ensuring optimal training for binary classification. |
| Optimizer | Adam | Adjusts learning rates dynamically for each parameter, leading to faster convergence and better training stability. |
| Epochs | 20 | The number of times the model passes through the entire dataset during training. More epochs improve learning but may cause overfitting. |

**Table 2:** Hyper parameters for initial binary classification CNN

To overcome the shortcomings of the initial model, a novel approach was developed, integrating Particle Swarm Optimization (PSO) for hyperparameter tuning. PSO has been proven effective in optimizing deep learning hyperparameters in medical imaging tasks by dynamically adjusting parameters for better convergence (Houssein et al., 2021). Instead of training a CNN from scratch, MobileNetV2, a pre-trained image classification model trained on ImageNet, was used as the base feature extractor, as shown by the teal blocks in figure 4. Studies have shown that transfer learning with pre-trained models significantly improves accuracy in histopathological image classification while reducing the need for extensive training data (Srinivasu et al., 2021). The dataset was divided based on magnifications (40X, 100X, 200X, 400X), ensuring that each model was trained on a consistent image resolution, a method that has been recommended for improving classification performance in multi-resolution medical imaging (Bayramoglu et al., 2017). To address class imbalance, data balancing was performed within each magnification level, maintaining an average of 250–350 images per class. However, given the limited dataset size, data augmentation techniques were implemented using ImageDataGenerator, applying transformations such as rescaling (1./255), width and height shifts (0.2), shear (0.2), zoom (0.2), and horizontal flipping to artificially increase the number of training samples. Studies indicate that data augmentation can significantly improve model generalization in medical imaging when training data is scarce (Bandi et al., 2018). PSO dynamically optimized the learning rate and dropout rate (represented by the dark blue blocks in figure 4), selecting values within the range of 1e-5 to 1e-2

(learning rate) and 0.3 to 0.7 (dropout rate), improving convergence and reducing overfitting, a technique that has been effectively used in medical image classification (Houssein et al., 2021). Unlike the previous approach, hyperparameter tuning was first conducted on 40X magnification, and once optimal parameters were determined, the model was trained separately on the remaining magnifications.

To further enhance training efficiency, ReduceLROnPlateau (factor 0.5, patience 3, min LR 1e-6) was introduced, dynamically adjusting the learning rate based on validation loss trends. Additionally, EarlyStopping (patience = 5, restore_best_weights = True) was implemented to halt training when improvements stagnated, preventing unnecessary computation and overfitting, an approach validated in previous deep learning studies (Srinivasu et al., 2021). Figure 4 below shows a diagrammatic representation of the CNN architecture used.

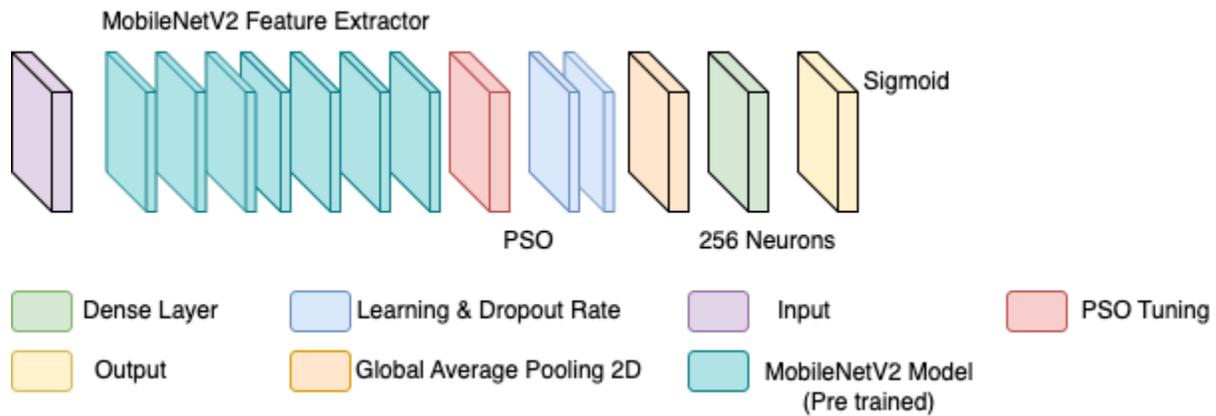

**Fig 4:** Updated CNN

However, this approach too yielded an accuracy far below acceptable levels. F1 scores, recall and precision also hovered around the 50% mark, forcing another change in approach.

| Hyperparameter | Value(s) Used | Explanation |
| --- | --- | --- |
| Optimizer | Particle Swarm Optimization (PSO) | Instead of using Adam, PSO dynamically selects the best learning rate and dropout rate by minimizing validation loss. This helps in fine-tuning model performance more efficiently. |
| Learning Rate | Optimized by PSO (Range: 1e-5 to 1e-2) | Controls the step size of weight updates. A smaller learning rate ensures stable learning, while a larger one speeds up convergence but risks overshooting the optimal point. |
| Dropout Rate | Optimized by PSO (Range: 0.3 to 0.7) | Dropout prevents overfitting by randomly deactivating neurons during training. The PSO algorithm selects the best dropout rate for optimal generalization. |
| Batch Size | 16 | The number of images processed at once during training. A smaller batch size helps conserve memory but may result in higher training variance. |
| Input Image Size | (128, 128, 3) | The input images are resized to 128×128 pixels with 3 color channels to ensure uniformity and reduce computational complexity. |

| Base Model | MobileNetV2 (Pre-trained on ImageNet) | A lightweight deep learning model optimized for mobile and embedded vision applications. The feature extraction layers are frozen to retain learned representations. |
|---|---|---|
| Magnification-Based Training | 40X, 100X, 200X, 400X (Trained Separately) | Instead of training all images together, the dataset is split based on magnification levels to improve feature learning at each resolution. |
| Data Augmentation | Rescaling (1./255), Width/Height Shift (0.2), Shear (0.2), Zoom (0.2), Horizontal Flip | Artificially increases dataset size by applying transformations, helping the model generalize better to unseen samples. |
| Validation Split | 20% | A portion of the dataset is reserved for validation to evaluate model performance and prevent overfitting. |
| Loss Function | Binary Crossentropy | Since the classification is binary (benign vs. malignant), binary crossentropy is used to compute the error between predicted and actual labels. |
| Activation Function | ReLU (Hidden Layers), Sigmoid (Output Layer) | ReLU helps prevent vanishing gradients in hidden layers, while Sigmoid is used in the output layer for binary classification (probabilities between 0 and 1). |
| ReduceLROnPlateau | Factor = 0.5, Patience = 3, Min LR = 1e-6 | Reduces the learning rate if validation loss stops improving for 3 consecutive epochs, preventing unnecessary weight updates. |
| EarlyStopping | Monitor = val_loss, Patience = 5, Restore Best Weights = True | Stops training if validation loss doesn't improve for 5 consecutive epochs, preventing overfitting and saving computation time. |
| Epochs | 30 | The number of times the model sees the entire dataset during training. Early stopping ensures the model doesn't train longer than necessary. |

**Table 3:** Hyperparameters for first attempt of per magnification training

The classification of breast cancer histopathology images involved separating the dataset based on different magnification levels (40X, 100X, 200X, and 400X). While this method aimed to leverage magnification-specific features, it resulted in severe dataset fragmentation. Each individual model was trained on a significantly smaller subset of images, which limited the ability to generalize across different test samples. Even with extensive data augmentation, the number of available images remained insufficient, leading to high variance and suboptimal performance during evaluation. Additionally, the previous models relied on relatively shallow convolutional neural networks (CNNs) for feature extraction. These networks struggled to effectively capture both low-level and high-level representations within the histopathology images, leading to limited discriminatory power when distinguishing between benign and malignant samples. Given these challenges, a more robust and generalized model was required to improve classification performance across all magnification levels To overcome the limitations of dataset fragmentation and poor feature extraction, a new model was developed using DenseNet121 as the backbone for feature extraction. DenseNet121, a deep convolutional neural network pre-trained on the ImageNet dataset, was chosen due to its efficient parameter utilization and ability to capture intricate hierarchical features. Unlike the previous approach, this model does not separate images based on magnification, allowing for a larger and more diverse training dataset, thereby improving generalization. The proposed model extracts feature from three different depths of DenseNet121: conv3_block12_concat, conv4_block24_concat, and conv5_block16_concat. These layers correspond to different levels of abstraction within the network, ensuring that both fine-grained and high-level morphological characteristics are captured. The selection of these layers allows for multi-scale feature extraction, improving the model's ability to differentiate between benign and malignant samples. The extracted features are processed through a structured pipeline to refine and enhance their representation before classification.

Feature extraction in this model occurred at three levels. The lowest-level feature map, obtained from conv3_block12_concat, captures fundamental visual characteristics such as edges, textures, and color distributions. These features are crucial for identifying initial structural differences in histopathology images, such as variations in cellular arrangements. The mid-level feature map, extracted from conv4_block24_concat, represents more complex patterns, including structural organization within the tissue and variations in gland formation. These features help differentiate between normal and abnormal cellular arrangements. The highest-level feature map, obtained from conv5_block16_concat, provides a more abstract representation, focusing on morphological changes indicative of malignancy, such as nuclear pleomorphism, mitotic activity, and stromal alterations. By incorporating multiple feature extraction levels, the model ensures a comprehensive understanding of tissue characteristics rather than relying on a single-layer representation. To ensure a balance between computational efficiency and adequate feature resolution, an input image size of 128×128 was chosen over 256×256. While higher resolutions like 256×256 could preserve more fine-grained tissue structures, they also significantly increase memory usage and computational load. Given that histopathology images already exhibit high variability and detailed cellular patterns, 128×128 provides a sufficient level of detail for feature extraction while allowing for larger batch sizes, faster training, and reduced GPU memory constraints. This choice is particularly important when training multiple models, as excessive computational demands could slow down hyperparameter tuning and optimization.

Each extracted feature map undergoes Global Average Pooling (GAP) to reduce dimensionality while retaining essential spatial information. Unlike max pooling, which selects only the most prominent activations, GAP ensures that all features contribute proportionally to the final representation, enhancing robustness and stability. After pooling, L2 normalization is applied to ensure numerical stability and prevent dominance of certain feature values due to large magnitudes. The normalized feature vectors are then transformed through fully connected layers, where a 64-neuron Dense layer introduces non-linearity, allowing for more complex feature interactions. Batch normalization follows to stabilize learning by standardizing feature distributions and improving gradient flow, leading to faster convergence and better generalization. This entire process for one feature is shown by the four blocks in each row in Figure 5. Instead of treating each extracted feature set independently, the model employs a fusion mechanism where feature representations from all three levels are concatenated into a unified feature descriptor (the white block in Figure 5). This approach integrates information across different abstraction levels, allowing the model to make more informed classification decisions. The fused feature vector undergoes further transformation through a Dense layer with 16 neurons, refining the representation before classification. To prevent overfitting, a Dropout layer with a probability of 0.45 is applied at this stage, randomly deactivating neurons during training and ensuring that the model does not rely on specific patterns that may not generalize well.

The final classification layer consists of a softmax activation function with two neurons, representing the benign and malignant classes. This output layer assigns a probability score to each class, allowing for precise decision-making in binary classification tasks. By leveraging deep feature extraction, multi-scale fusion, and regularization techniques, the proposed model significantly improves the accuracy and robustness of breast cancer classification in histopathology images as compared to the previous 2 as well as objectively. Figure 4 diagrammatically describes the flow of the updated CNN architecture.

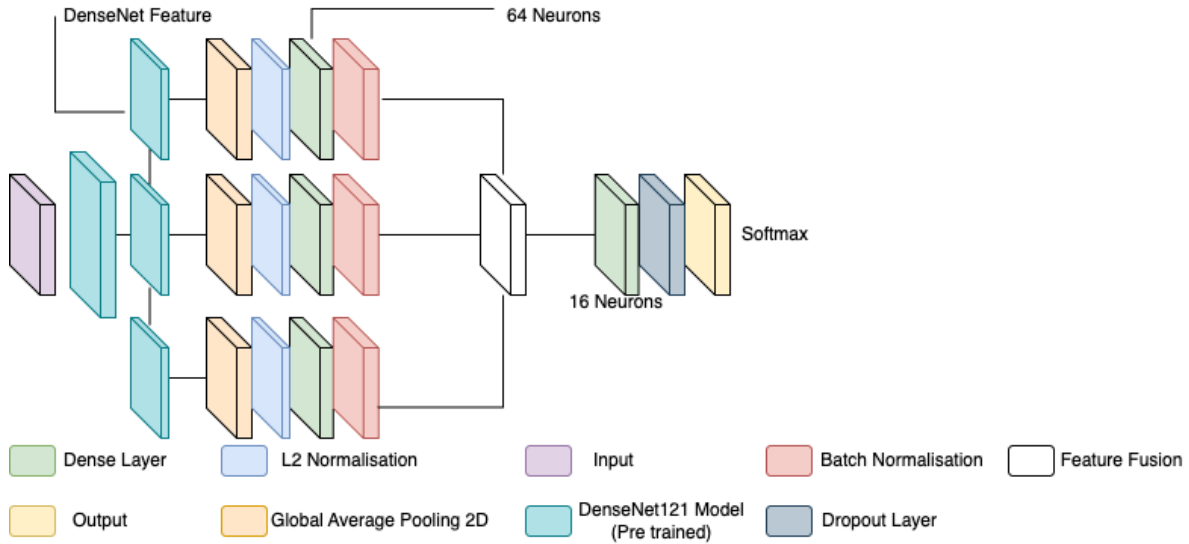

**Fig 5:** Final Binary Class CNN

| Hyperparameter | Value | Explanation |
| --- | --- | --- |
| Base Model | DenseNet121 (pre-trained on ImageNet) | Acts as the backbone for feature extraction, leveraging pre-trained hierarchical features. |
| Input Image Size | (128, 128, 3) | Defines the input shape of images with 3 color channels (RGB), ensuring consistency across the dataset. |
| Feature Extraction Layers | conv3_block12_concat, conv4_block24_concat, conv5_block16_concat | Extracts multi-scale features from different levels of abstraction within the DenseNet121 model. |
| Pooling Method | Global Average Pooling (GAP) | Reduces dimensionality while retaining essential spatial information from feature maps. |
| Feature Normalization | L2 Normalization | Ensures all extracted features contribute proportionally, preventing large-magnitude features from dominating learning. |
| Hidden Dense Layer (Feature Processing) | 64 neurons, ReLU activation, L2 regularization (0.001) | Introduces non-linearity to enhance feature interactions and prevents overfitting via L2 regularization. |
| Batch Normalization | Applied after Dense layers | Standardizes feature distributions, improving gradient flow and training stability. |
| Feature Fusion | Concatenation of three feature levels | Integrates low-level, mid-level, and high-level features into a unified representation for better classification. |
| Final Dense Layer | 16 neurons, ReLU activation, L2 regularization (0.001) | Further refines the fused feature representation before classification. |
| Dropout Rate | 0.45 | Randomly deactivated neurons to prevent overfitting, ensuring better generalization. |
| Output Layer | 2 neurons, Softmax activation | Generates class probabilities for binary classification (Benign vs. Malignant). |
| Loss Function | Categorical Crossentropy | Suitable for multi-class classification tasks, even though the output is binary (ensures numerical stability). |
| Optimizer | Adam (learning rate: 0.0001) | Adaptive optimization algorithm that adjusts learning rates dynamically for efficient training. |
| Learning Rate Scheduler | ReduceLROnPlateau (factor=0.5, patience=3, min_lr=1e-6) | Reduces the learning rate when validation loss stagnates, helping fine-tune the model. |
| Early Stopping | Patience = 5, restore_best_weights = True | Stops training when validation loss stops improving, preventing unnecessary computation and overfitting. |
| Epochs | 50 | Maximum number of iterations through the dataset, ensuring the model learns enough before early stopping. |

**Table 4:** Hyperparameters for the final binary classification model

### 2.2.2. Detecting between the multiple subclasses

There are four histological distinct types of benign breast tumors: adenosis (A), fibroadenoma (F), phyllodes tumor (PT), and tubular adenoma (TA); and four malignant tumors (breast cancer): carcinoma (DC), lobular carcinoma (LC), mucinous carcinoma (MC) and papillary carcinoma (PC).

Adenosis (A) is characterized by the proliferation of glandular structures, which may appear as distorted acini or ducts in biopsy images. These lesions typically show increased glandular tissue without significant atypia, making them relatively easy to differentiate from malignancies under a microscope (Jiang et al., 2017). Fibroadenomas (F) are solid tumors composed of both glandular and stromal components, presenting as well-circumscribed masses with minimal cellular atypia. Their typical pattern in biopsy images shows a lobular architecture with a fibrous stroma, which contrasts with the irregular structures seen in malignant lesions (Bani et al., 2015). Phyllodes tumors (PT), though benign in some cases, can show features suggestive of malignancy, such as rapid growth and hypercellularity. Their biopsy images typically display leaf-like structures, hence the name, with prominent stromal overgrowth and areas of necrosis that can challenge differentiation from malignant tumors (Barker et al., 2018). Tubular adenomas (TA) are another benign subtype characterized by well-formed tubules lined by a single layer of epithelial cells. These lesions often have minimal cellular atypia, and their biopsy images demonstrate small, uniformly sized tubules with little stroma (Pike et al., 2020).

In contrast, malignant breast cancers are often more challenging to classify due to the diversity of their histopathological features. Ductal carcinoma (DC) is the most common type of breast cancer and typically appears as irregularly shaped masses with infiltrating ductal structures. The biopsy images show marked cellular pleomorphism, nuclear atypia, and often necrotic areas, which are indicative of aggressive behavior (Tung et al., 2016). Lobular carcinoma (LC) presents with a distinctive pattern in biopsy images, with tumor cells growing in a single-file pattern and a lack of cohesive cell groups. This "Indian file" arrangement of cells and the absence of desmoplastic stroma are key features that differentiate it from ductal carcinoma (Vargas et al., 2017). Mucinous carcinoma (MC) is characterized by the production of extracellular mucin, which can be observed in the biopsy images as abundant mucin pools, separating the tumor cells. The cells in mucinous carcinoma are typically round with mild pleomorphism and are surrounded by mucin-rich stroma (Manning et al., 2018). Finally, papillary carcinoma (PC) shows prominent papillary structures with fibrovascular cores, surrounded by atypical epithelial cells. The tumor cells are often arranged in well-defined papillae, and the biopsy images reveal a complex architecture with cystic spaces and dense fibrous stroma (Thompson et al., 2019).

A Convolutional Neural Network has been used to differentiate between these subtypes by leveraging its ability to learn hierarchical spatial features from biopsy images. In particular, the CNN has been trained to recognize the subtle differences in architectural patterns, such as the well-defined tubules in tubular adenomas versus the irregular glandular structures in ductal carcinoma. The network was also trained to identify key histological features such as cellular pleomorphism, stromal composition, and the presence of mucin, which are crucial for distinguishing between different tumor types (Rajendran et al., 2018). By extracting features like texture, shape, and size from biopsy images, a CNN model has been fine-tuned to classify benign and malignant tumors with high accuracy, even differentiating between subtypes of cancer like mucinous carcinoma and papillary carcinoma based on their unique histological patterns (Srinivasan et al., 2020).

The selected hyperparameters in table 5 play a crucial role in optimizing the performance of the convolutional neural network for breast cancer histology classification. The input image size of (256,256) ensures that all images maintain a uniform shape, allowing the network to process them efficiently. The batch size of 32 determines how many images are processed in one training step, balancing memory efficiency and training speed. Different learning rates for Adam (0.0001), SGD (0.01), and RMSprop (0.0001) affect how quickly the model updates its weights; a smaller learning rate helps prevent overshooting the optimal weights, while a larger one speeds up convergence. Additionally, SGD momentum (0.9) stabilizes updates by reducing oscillations, leading to more effective training. The architecture of the model relies on three Conv2D layers with 32, 64, and 128 filters, allowing the network to extract increasingly complex features. A (3,3) kernel size is used for convolution operations, ensuring effective feature extraction without excessive computational cost. The ReLU activation function introduces non-linearity, allowing the network to learn complex patterns. Dropout rates of 0.25 for convolutional layers and 0.5 for dense layers prevent overfitting by randomly deactivating neurons during training. The (2,2) pooling size reduces feature map dimensions, enhancing computational efficiency. The dense layer with 512 neurons serves as a bridge between convolutional layers and the final classification. Finally, the softmax activation function in the output layer ensures proper probability distribution for multi-class classification, while the categorical crossentropy loss function effectively measures classification errors. Training for 30 epochs allows the model to learn effectively without excessive overfitting. However, as was the case for the initial two approaches of the binary classification model, this too yielded poor results. Hence, after the binary model gave satisfactory results, it was further extended to the multiclass version of the same.

| Hyperparameter | Value | Explanation |
| --- | --- | --- |
| Input Image Size | (256, 256) | The spatial dimensions of each input image fed to the model. |
| Batch Size | 32 | Number of images processed at once during training. |
| Learning Rate (Adam) | 0.0001 | Speed at which the model updates weights during training using Adam optimizer. |
| Learning Rate (SGD) | 0.01 | Speed of weight updates for the SGD optimizer. |
| SGD Momentum | 0.9 | Momentum helps accelerate SGD by dampening oscillations during updates. |
| Learning Rate (RMSprop) | 0.0001 | Step size for weight updates using RMSprop optimizer. |
| Conv2D Filters (1st Layer) | 32 | Number of filters in the first convolutional layer. |
| Conv2D Filters (2nd Layer) | 64 | Number of filters in the second convolutional layer. |
| Conv2D Filters (3rd Layer) | 128 | Number of filters in the third convolutional layer. |
| Kernel Size | (3, 3) | Size of the sliding window applied in convolution operations. |

| Activation Function | ReLU | Non-linear function applied to neuron outputs to introduce non-linearity. |
|---|---|---|
| Dropout Rate (Conv Layers) | 0.25 | The fraction of neurons dropped to prevent overfitting during training. |
| Dropout Rate (Dense Layer) | 0.5 | Fraction of dense layer neurons dropped for regularization. |
| Pooling Size | (2, 2) | Size of the window for max pooling operation to reduce feature map dimensions. |
| Dense Layer Size | 512 | Number of neurons in the fully connected layer before the output. |
| Output Layer Activation | Softmax | Ensures output values represent probabilities for multi-class classification. |
| Loss Function | Categorical Crossentropy | Measures model error for multi-class classification tasks. |
| Number of Epochs | 30 | Total number of complete passes through the training dataset. |

**Table 5:** Hyperparameters for the initial subclass classification model

To extend the binary classification framework to a multi-class setting, two separate CNNs were developed—one for classifying benign subtypes and another for malignant subtypes. This approach ensures that each network specializes in distinguishing fine-grained variations within its respective category, leading to more precise feature extraction and improved classification accuracy. The multi-class CNNs are built upon the binary classification architecture, using DenseNet121 as the backbone for deep feature extraction. DenseNet121, pre-trained on the ImageNet dataset, was selected due to its ability to efficiently capture complex hierarchical features while maintaining parameter efficiency through dense connectivity. Unlike traditional CNNs, DenseNet121 propagates feature maps from earlier layers directly to deeper layers, facilitating multi-scale learning and improved gradient flow. Feature extraction occurs at three critical depths of DenseNet121: conv3_block12_concat, conv4_block24_concat, and conv5_block16_concat. These layers correspond to different levels of abstraction, ensuring that the model captures a comprehensive range of morphological features. The first extracted feature map, obtained from conv3_block12_concat, represents fundamental visual patterns such as edges, textures, and basic structural formations within the tissue. These low-level features help identify early distinctions in cellular architecture. The second feature map, extracted from conv4_block24_concat, captures more complex organizational structures, such as glandular formations and stromal variations, which are critical for differentiating subtypes within benign and malignant groups. The highest-level feature map, obtained from *conv5_block16_concat*, encodes abstract morphological patterns, including nuclear pleomorphism, mitotic activity, and other high-level histopathological changes associated with cancer progression.

Once extracted, these feature maps undergo Global Average Pooling (GAP) to reduce dimensionality while preserving essential spatial characteristics. GAP ensures that the model retains information from the entire feature map rather than focusing solely on the most dominant activations, as would occur with max pooling. This results in more stable feature representations and reduces sensitivity to localized artifacts. Following GAP, the feature vectors undergo L2 normalization, which ensures that all extracted features contribute proportionally to the classification process by preventing large-magnitude values from dominating the learning process. Each normalized feature vector is passed through an independent Dense layer with 64 neurons, introducing non-linearity and enhancing feature transformation. To prevent overfitting and encourage better generalization, L2 regularization is applied to these Dense layers. Next, Batch Normalization is employed to standardize feature distributions, improving gradient flow and accelerating convergence during training. The process the features undergo is represented in Figure 7 as a flowchart. These processed feature representations from all three levels are then concatenated into a unified feature vector, allowing the model to integrate fine-grained, mid-level, and high-level information into a single, multi-scale representation. The fused feature vector is further refined through an additional Dense layer with 16 neurons, reducing redundancy and improving the discriminative power of the final representation. To mitigate overfitting, a Dropout layer with a probability of 0.45 is introduced, randomly deactivating neurons during training to prevent reliance on specific patterns that may not generalize well. The final classification is performed through a softmax activation layer with four neurons, corresponding to the four subtypes within either the benign or malignant category. This softmax layer assigns a probability score to each subtype, allowing the model to make precise multi-class predictions.

The network is optimized using the Adam optimizer with a learning rate of 0.0001, ensuring effective weight updates while maintaining training stability. To further improve generalization and prevent overfitting, Early Stopping is employed, terminating training when validation loss ceases to improve. Additionally, ReduceLROnPlateau dynamically adjusts the learning rate by reducing it when validation loss stagnates, allowing the model to fine-tune its learning process for better convergence. The decision to develop separate CNNs for benign and malignant classifications was motivated by the distinct morphological characteristics of each category. A single CNN trained on all subtypes might struggle with feature entanglement, where overlapping patterns between benign and malignant samples obscure important differences. By training separate networks, each CNN focuses exclusively on intra-class variations, enabling more precise learning of subtype-specific structures. This specialization improves the ability to distinguish between fine-grained patterns within each category, ultimately enhancing the accuracy and reliability of breast cancer subtype classification. Figure 5 below shows the multiclass CNN as a diagram. Figure 6 shows the process each DenseNet feature goes through before being fused into one feature vector.

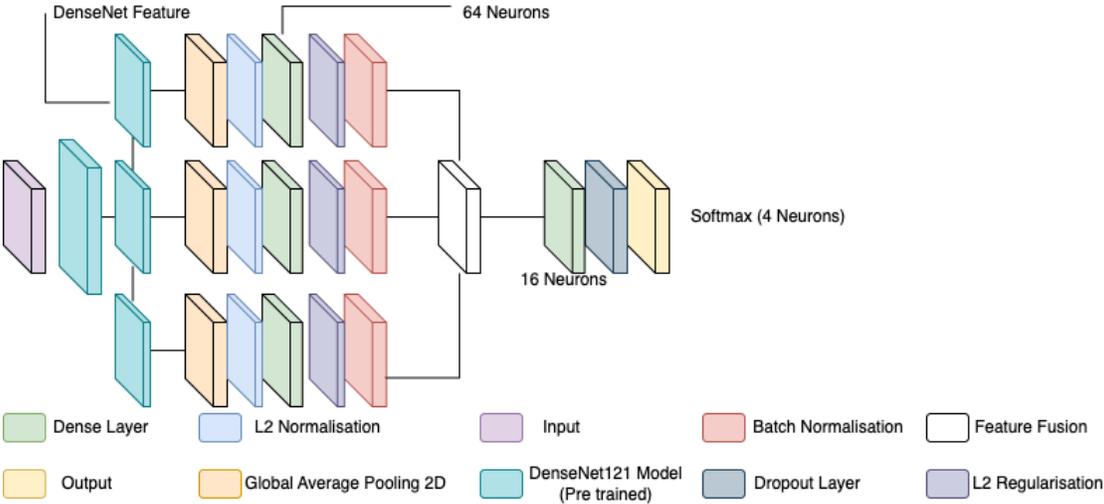

**Fig 6:** CNN used for Multi Class Classification

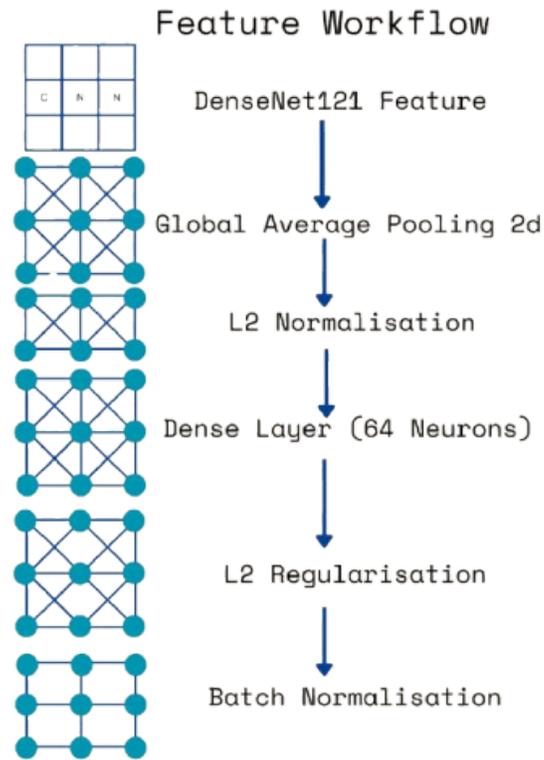

**Fig 7:** Diagram for Feature Workflows

| Hyperparameter | Value | Explanation |
| --- | --- | --- |
| Base Model | DenseNet121 (pre-trained on ImageNet) | Acts as a deep feature extractor, leveraging hierarchical feature learning for efficient classification. |
| Input Image Size | (128, 128, 3) | Ensures computational efficiency while preserving essential histopathological details. A higher resolution like 256×256 would increase memory usage without significantly improving feature representation. |
| Feature Extraction Layers | conv3_block12_concat, conv4_block24_concat, conv5_block16_concat | Extracts features at multiple abstraction levels, capturing both fine-grained and high-level morphological details. |
| Pooling Method | Global Average Pooling (GAP) | Reduces dimensionality while preserving spatial characteristics of the extracted feature maps. |
| Feature Normalization | L2 Normalization | Prevents feature values with high magnitudes from dominating the learning process, ensuring numerical stability. |
| Hidden Dense Layer (Feature Processing) | 64 neurons, ReLU activation, L2 regularization (0.001) | Introduces non-linearity for complex feature interactions and applies L2 regularization to prevent overfitting. |
| Batch Normalization | Applied after Dense layers | Standardizes feature distributions, accelerating convergence and improving training stability. |
| Feature Fusion | Concatenation of three feature levels | Integrates information across different levels of abstraction, enhancing classification performance. |
| Final Dense Layer | 16 neurons, ReLU activation, L2 regularization (0.001) | Refines the multi-scale feature representation before classification. |
| Dropout Rate | 0.45 | Prevents overfitting by randomly deactivating neurons, ensuring better generalization. |
| Output Layer | 4 neurons, Softmax activation | Outputs probability scores for the four benign subtypes. |
| Loss Function | Categorical Crossentropy | Suitable for multi-class classification tasks, optimizing probability-based learning. |
| Optimizer | Adam (learning rate: 0.0001) | Adaptive optimization that adjusts learning rates dynamically for efficient and stable training. |
| Learning Rate Scheduler | ReduceLROnPlateau (factor=0.5, patience=3, min_lr=1e-6) | Reduces learning rate when validation loss stagnates, improving fine-tuning of weights. |
| Early Stopping | Patience = 5, restore_best_weights = True | Stops training when validation loss stops improving, preventing unnecessary computation and overfitting. |
| Epochs | 50 | Maximum number of training iterations through the dataset, ensuring sufficient learning. |

**Table 6:** Hyperparameters for the final subclass classification model

## 2.3 Metrics Evaluations

In artificial intelligence, especially in domains like classification and regression, evaluation metrics play a vital role in measuring model performance. Within medical diagnostics, metrics such as precision and recall are particularly significant. Precision reflects the proportion of correct positive predictions, meaning that when a model identifies cancer, it is likely to be correct—thereby reducing unnecessary alarm (Sokolova & Lapalme, 2009). Recall, on the other hand, focuses on identifying as many actual positive cases as possible, which is crucial in healthcare settings to avoid overlooking genuine cases of disease (Sokolova & Lapalme, 2009). Although accuracy is a common metric indicating the percentage of correctly predicted instances overall, it can be deceptive in scenarios with class imbalance, as it does not account for the nature of errors made (Tharwat, 2020).

To address such limitations, the F1-score is often employed. This metric, calculated as the harmonic mean of precision and recall, offers a more balanced view by considering both false positives and false negatives—an essential factor in medical imaging, where failing to detect malignant tumors can have serious consequences (Chicco & Jurman, 2020). Additionally, the Area Under the Curve (AUC) of the Receiver Operating Characteristic (ROC) curve is another key metric. It evaluates the model's capacity to distinguish between positive (cancerous) and negative (non-cancerous) cases, with scores approaching 1 indicating high discriminative ability (Yang & Ying, 2022). Given the frequent imbalance in medical datasets, relying on a combination of these metrics provides a more reliable and holistic evaluation of model effectiveness.

| Metric | Formula | Explanation | References |
|---|---|---|---|
| Precision | TP/(TP+FP) | Measures the proportion of correctly predicted positive cases out of all predicted positives. | Sokolova & Lapalme, 2009 |
| Recall (Sensitivity) | TP/(TP+FN) | Measures the proportion of actual positive cases correctly identified by the model. | Sokolova & Lapalme, 2009 |
| F1-Score | 2(Precision*Recall)/(Precision+Recall) | Harmonic mean of precision and recall, providing a balanced measure for imbalanced datasets. | Chicco & Jurman, 2020 |
| Accuracy | (TP+TN)/(TP+TN+FP+FN) | Measures overall correctness of the model across all classes. | Tharwat, 2020 |
| AUC (Area Under the Curve) | Computed from ROC curve | Represents the probability that the model ranks a randomly chosen positive instance higher than a randomly chosen negative one. | Yang & Ying, 2022 |

**Table 7:** Metrics for evaluation of algorithm

- ❖ *TP (True Positives): Correctly predicted positive cases.*
- ❖ *TN (True Negatives): Correctly predicted negative cases.*
- ❖ *FP (False Positives): Incorrectly predicted positive cases.*
- ❖ *FN (False Negatives): Incorrectly predicted negative cases.*

## 3. Results & discussion

The following sections present a comprehensive evaluation of the developed AI models across various experimental phases, including binary classification and multi-class subtype detection for breast cancer biopsy images. Detailed performance metrics such as accuracy, precision, recall, F1-score, and ROC-AUC are reported to assess the effectiveness and generalizability of each model. The study began by analyzing the outcomes of the initial CNN architecture, followed by performance improvements observed with the final DenseNet121-based model. It then extends this evaluation to multi-class classification tasks, detailing the results for both benign and malignant subtypes. These results collectively demonstrate the diagnostic power and clinical viability of the proposed framework.

### 3.1 The Binary Categorical Classification of the 2 classes- Malignant & Benign

The initial CNN architecture, developed for binary classification of histopathology images into benign and malignant categories, was evaluated using different filter and kernel size configurations. The table presented in this section summarizes the performance metrics for each configuration. While the model achieved a maximum accuracy of 84%, this result was below the expected benchmark for binary classification tasks, where an accuracy of 95% or higher is generally required to maintain robust performance during the transition to multi-class classification.

| Filters | Mesh Size | Precision | Recall | Binary Accuracy | AUC |
|---|---|---|---|---|---|
| 32, 32, 32 | 3x3 | 0.8603 | 0.9283 | 0.8424 | 0.8768 |
| 16, 32, 16 | 4x4 | 0.8893 | 0.8994 | 0.8542 | 0.8872 |
| 16, 32, 16 | 5x5 | 0.8253 | 0.8465 | 0.7799 | 0.8290 |

Table 8: Initial Results of Binary classification

Upon using the proposed DenseNet121-based binary classification, the model demonstrated outstanding performance in distinguishing between benign and malignant breast cancer histopathology images. The model was trained on out of sample biopsy images. As shown in Table 9, the model achieved a test accuracy of 98.63%, indicating a high level of reliability in classification. Precision, recall, and F1-score values were also notably high at 0.9888, 0.9867, and 0.9877, respectively, reflecting the model's ability to correctly identify malignant cases while minimizing false positives and false negatives. The ROC-AUC score of 0.9863 further confirms the model's strong discriminative capability, showing its robustness in distinguishing between the two classes across different classification thresholds. Figure 8 shows the confusion matrix for the classification while Figure 9 is a graph showing the change in accuracy and loss with respect to epochs. Compared to previous models that relied on shallow convolutional architectures and dataset fragmentation based on magnification levels, the proposed model benefits from multi-scale feature extraction at different network depths, improving its ability to generalize across varying histopathological patterns. By utilizing a 128×128 input image resolution, the model strikes a balance between computational efficiency and feature preservation, allowing for optimal learning without excessive memory consumption. The use of feature fusion across different abstraction levels enhances the classification robustness, ensuring that both low-level structural features and high-level morphological variations contribute to the final decision-making process.

| Metric | Value |
|---|---|
| Test Accuracy | 0.9863 |
| Precision | 0.9888 |
| Recall | 0.9867 |
| F1 Score | 0.9877 |
| ROC-AUC | 0.9863 |

Table 9: Final Results of Binary Classification algorithm

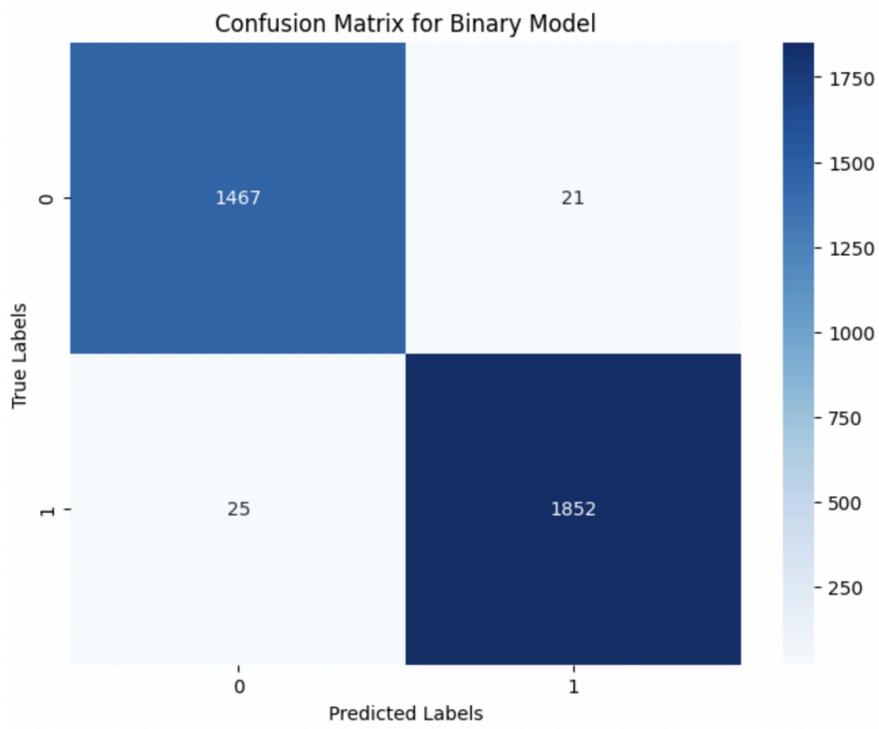

**Fig 8:** Confusion Matrix for Final Binary Classification

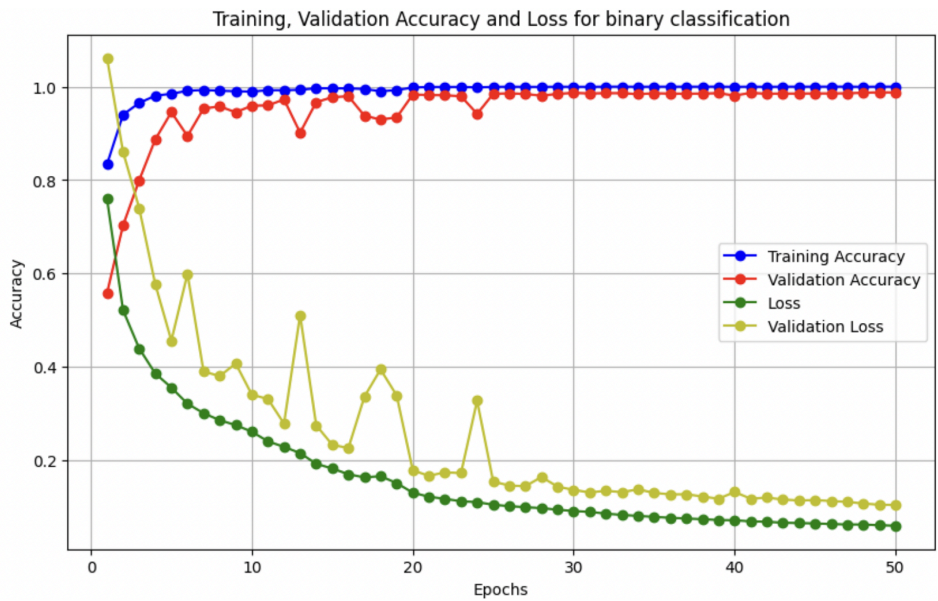

**Fig 9:** Accuracy/Loss vs Epoch Graph

**3.2 Multi Category Classification of the subclasses**

The performance of the multi-class classification models for benign and malignant breast cancer subtypes is summarized in Table 10. The model was trained on out of sample biopsy images. The model trained for benign classification achieved a test accuracy of 0.9382, whereas the malignant classification model obtained a test accuracy of 0.9158. Additionally, the malignant model demonstrated strong predictive capabilities, with a precision of 0.9151, recall of 0.9240, and an F1-score of 0.9182 while the benign model had a precision of 0.9541, recall of 0.9324 and an F1 score of 0.9415. These results highlight the effectiveness of using separate convolutional neural networks (CNNs) for each subtype, allowing for improved feature extraction and discrimination between fine-grained morphological variations. Figure 10 shows the confusion matrix for the benign model while figure 12 shows the same for the malignant model. Figure 11 shows the graph of the change in accuracy/loss with respect to the epochs for benign and figure 12 shows the same for malignant.

| Classification Task | Accuracy | Precision | Recall | F1 Score |
|---|---|---|---|---|
| Benign | 0.9483 | 0.9541 | 0.9324 | 0.9415 |
| Malignant | 0.9254 | 0.9318 | 0.9193 | 0.9251 |

**Table 10:** Final Results for Multi Category Classification

The results indicate that the model for benign subtypes outperformed the malignant classification model in terms of overall accuracy. However, the malignant classification model maintained strong recall and F1-score values, demonstrating its ability to correctly identify malignant subtypes while balancing precision. These findings support the hypothesis that training separate models for benign and malignant subtypes allows for more specialized learning, leading to improved classification performance. The integration of DenseNet121 as a feature extractor has further contributed to the model's ability to capture multi-scale morphological patterns in histopathology images.

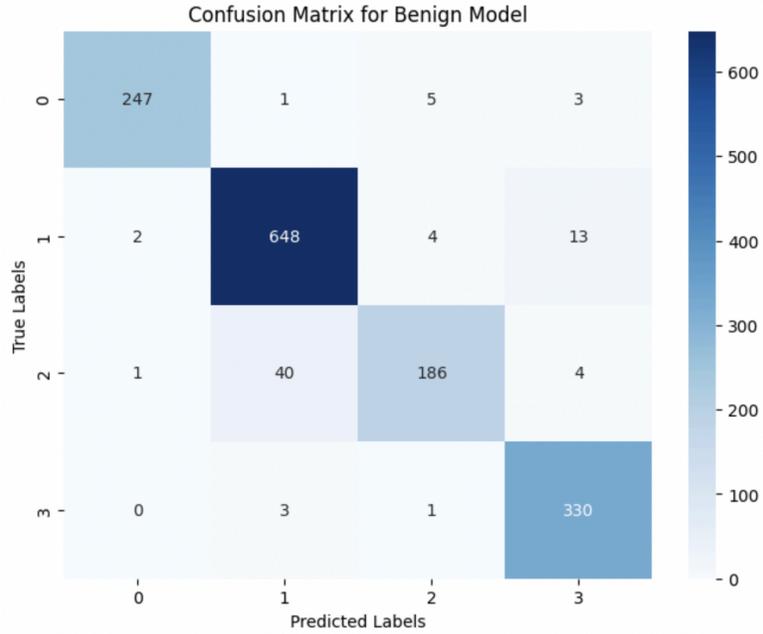

**Fig 10:** Confusion Matrix for Benign Sub Category Classification

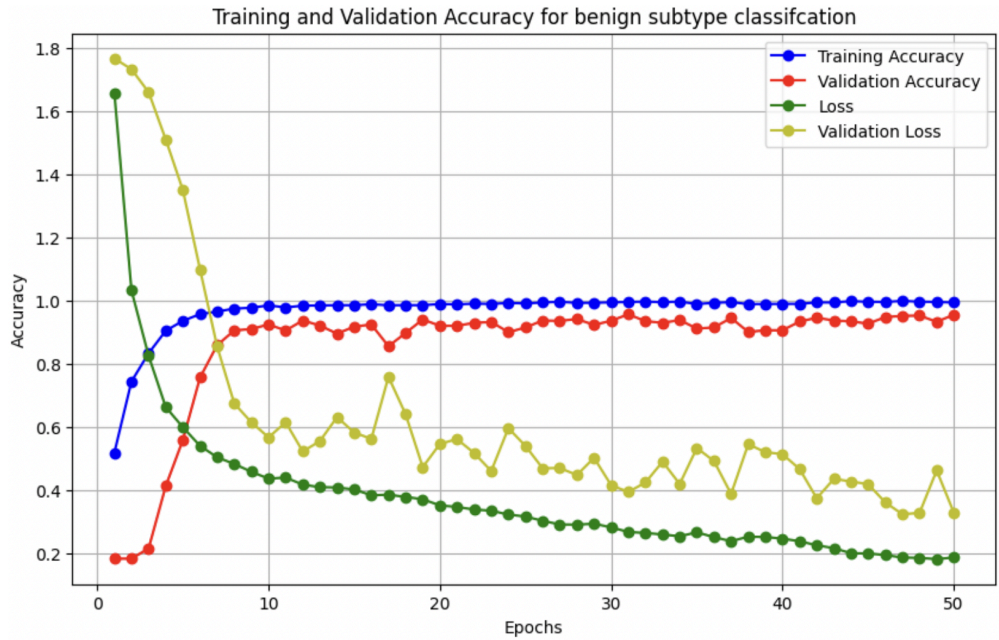

**Fig 11:** Accuracy/Loss vs Epoch Graph for Benign Subclass Model

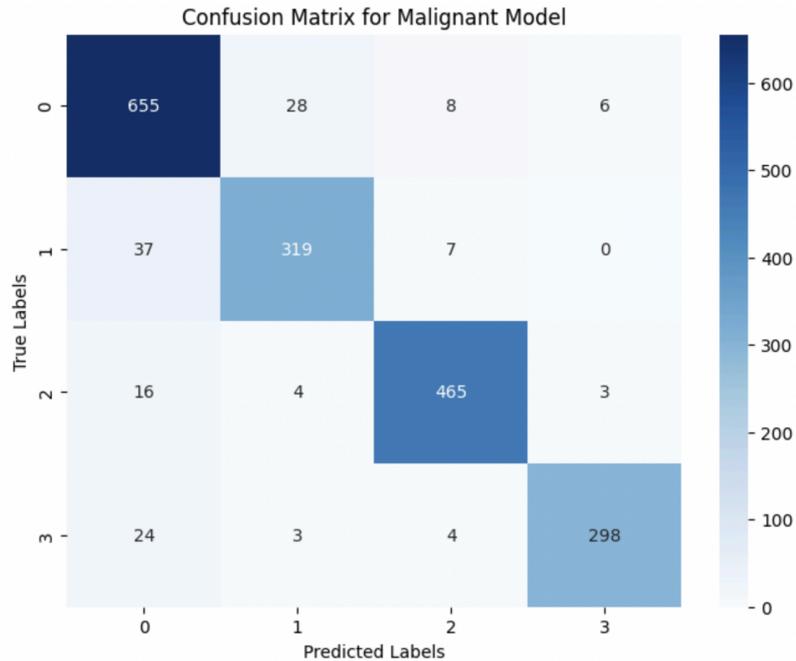

**Fig 12:** Confusion Matrix for Malignant Sub Category Classification

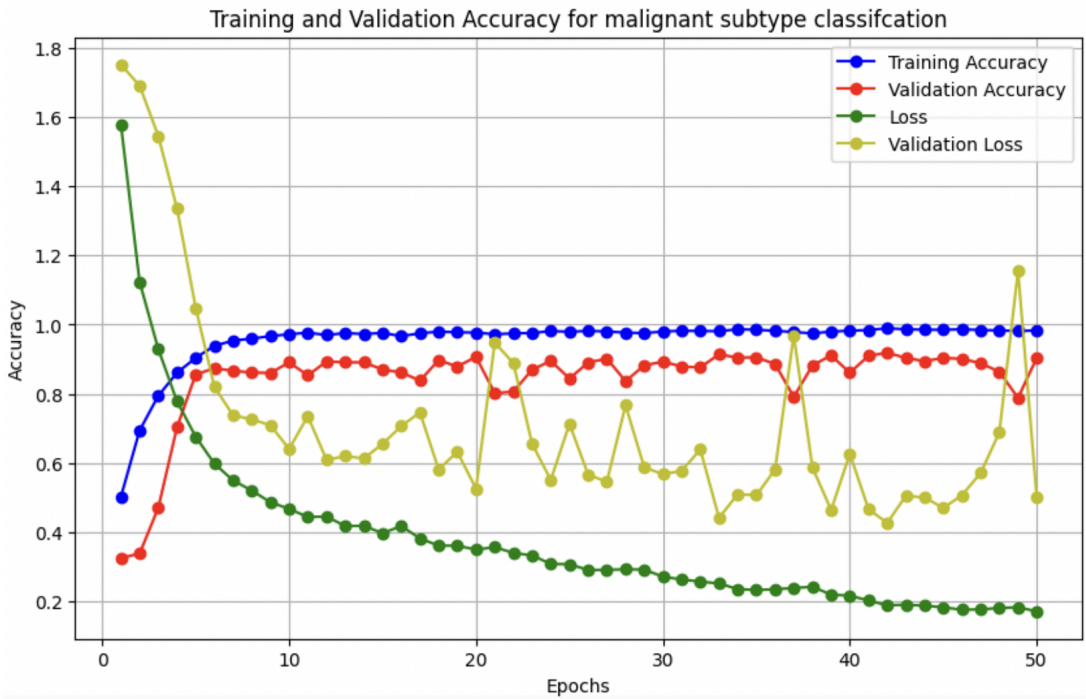

**Fig 13:** Accuracy/Loss vs Epoch Graph for Malignant Subclass Model

The final model (DenseNet feature extraction) demonstrated the highest performance among all approaches, achieving superior accuracy, recall, and F1 scores in binary classification of breast cancer biopsy images. By leveraging DenseNet121 with multi-scale feature extraction from three hierarchical layers, the model captured both low-level and high-level morphological patterns. Each feature map was processed through pooling, normalization, and dense layers before being fused into a unified representation. This architecture enabled precise and robust classification, effectively addressing the challenges of dataset fragmentation and limited feature representation seen in earlier models.

This algorithm lays the foundation for a future clinical tool that could revolutionize breast cancer diagnosis through histopathological biopsy analysis. In its envisioned form, the tool would allow users—such as pathologists or clinicians—to simply upload a biopsy image, upon which the AI system would process the image within minutes to determine whether the tissue is benign or malignant, and further classify it into one of the specific histological subtypes.

## 4. Conclusions

Clinically, this tool holds profound significance for the diagnosis and treatment planning of breast cancer, especially in its ability to perform subtype-level classification directly from biopsy images. Traditional diagnostic workflows often require multiple follow-up procedures beyond the initial biopsy—such as immunohistochemistry, molecular testing, and repeated imaging—to conclusively determine the histological subtype of a breast tumor (Robbins et al., 2010; National Cancer Institute, 2021; Manning et al., 2018). These additional tests are not only invasive but also delay the initiation of treatment and increase patient burden. In contrast, the proposed AI model achieves direct multi-class subtype classification from a single biopsy image using a dual-branch DenseNet121 architecture, significantly reducing diagnostic latency. Unlike other state-of-the-art models that typically achieve high accuracy in binary classification (e.g., 96.5% using DenseNet121 as reported by Zhang et al., 2024) or require external interpretability layers to explain decisions (as proposed by Lee et al., 2024), this tool uniquely integrates multi-scale feature fusion (from conv3_block12_concat, conv4_block24_concat, and conv5_block16_concat), L2 normalization, batch normalization, and dropout regularization into a unified, end-to-end framework. No model currently in the literature combines this level of architectural sophistication with interpretability and clinical applicability across both benign and malignant subtype classifications—making this tool a true first-of-its-kind in histopathological breast cancer diagnosis. Studies on AI-based tools for subtype classification have typically lacked this depth or specialization, focusing either on feature reuse or singular magnification inputs without tackling dataset fragmentation or multi-class challenges simultaneously (Patel et al., 2024).

This innovation does not just advance computational performance; it stands to redefine standards of care in oncology. By making it possible to obtain subtype-specific insights from a single excisional biopsy, this AI model could reduce dependence on additional molecular assays and eliminate the need for serial diagnostic interventions (Manning et al., 2018; National Cancer Institute, 2021). This is particularly impactful for women in low-resource settings, where access to specialized pathology services is limited and time to treatment initiation is often dangerously delayed (World Health Organization, 2021). The model's high recall and F1-scores across both benign (F1: 0.9415) and malignant subtypes (F1: 0.9251) ensure that clinically relevant distinctions—such as between phyllodes tumors and fibroadenomas, or mucinous and papillary carcinomas—can be made with precision. In doing so, this tool directly supports early, accurate, and non-redundant therapeutic planning, sparing patients from the emotional, financial, and physical toll of diagnostic uncertainty. With scalability, reproducibility, and interpretability at its core, this AI system represents a quantum leap in precision pathology, bringing us closer to the vision of fully personalized, efficient breast cancer care.